\documentclass[aps,prl,twocolumn,superscriptaddress,showpacs]{revtex4-1}
\usepackage{graphicx}
\usepackage{calc}
\usepackage{bm}

\begin{document}
 
\title{Energy transport by neutral collective excitations at the quantum Hall edge}

\author{E.V.~Deviatov}
\email[Corresponding author. E-mail:~]{dev@issp.ac.ru}
 \affiliation{Institute of Solid State
Physics RAS, Chernogolovka, Moscow District, 142432, Russia}

\author{A.~Lorke}
\affiliation{Laboratorium f\"ur Festk\"orperphysik, Universit\"at
Duisburg-Essen, Lotharstr. 1, D-47048, Duisburg, Germany}

\author{G.~Biasiol}
\affiliation{IOM CNR, Laboratorio TASC, 34149 Trieste, Italy}

\author{L.~Sorba}
\affiliation{NEST, Istituto Nanoscienze-CNR and Scuola Normale Superiore, 56127 Pisa, 
Italy}

\date{\today}

\begin{abstract}
We use the edge of the quantum Hall sample to study the possibility for counter-propagating neutral collective excitations. A novel sample design allows us to independently investigate charge and energy transport along the edge. We experimentally observe an upstream energy transfer with respect to the electron drift  for the filling factors 1 and 1/3. Our analysis indicates that a neutral collective mode at the interaction-reconstructed edge is a proper candidate for the experimentally observed effect.
\end{abstract}


\maketitle

Since the physics of quantum Hall (QH) effect is largely investigated now, QH regime can be conveniently used to model complicated phenomena from different areas of modern physics. Particularly, gapless collective excitations are of special interest in graphene physics, topological insulators, and quantum computation. In the QH regime the universal low-energy physics  is only connected with edge collective modes~\cite{wen}. It was proposed~\cite{kane}, that at some fractional filling factors interaction can lead to an unusual neutral collective mode, which propagates upstream with respect to electron drift and carries only energy. 

Neutral modes have not been observed in a direct heat-transport experiment~\cite{granger}, however, they were detected in  shot-noise measurements~\cite{heiblum}. This discrepancy might originate from different experimental methods, so an independent investigation in a different experimental configuration  is of great interest. 

Here, we use the reconstructed edge of the quantum Hall sample to study the possibility of counter-propagating neutral collective excitations. The edge reconstruction is predicted~\cite{yang} to result from the interplay between the smooth edge potential and the Coulomb interaction energy. Experimental  arguments for the reconstruction of this type can be found in the capacitance measurements at the edges of $\nu=1,1/3$ plateau, where a so-called negative compressibility was experimentally observed~\cite{eisen}. A novel sample design allows us to independently investigate charge and energy transport along the  edge. We experimentally observe an upstream energy transfer with respect to the electron drift  for the filling factors 1 and 1/3. Our analysis indicates that a neutral collective mode at the interaction-reconstructed edge is a proper candidate for the experimentally observed effect.

Our samples are fabricated from a molecular beam epitaxially-grown GaAs/AlGaAs heterostructure. It contains a two-dimensional electron gas (2DEG) located 200~nm below the surface. The 2DEG mobility at 4K is  $5.5 \cdot 10^{6}  $cm$^{2}$/Vs  and the carrier density is   $1.43 \cdot 10^{11}  $cm$^{-2}$.

\begin{figure}
\includegraphics*[width=0.95\columnwidth]{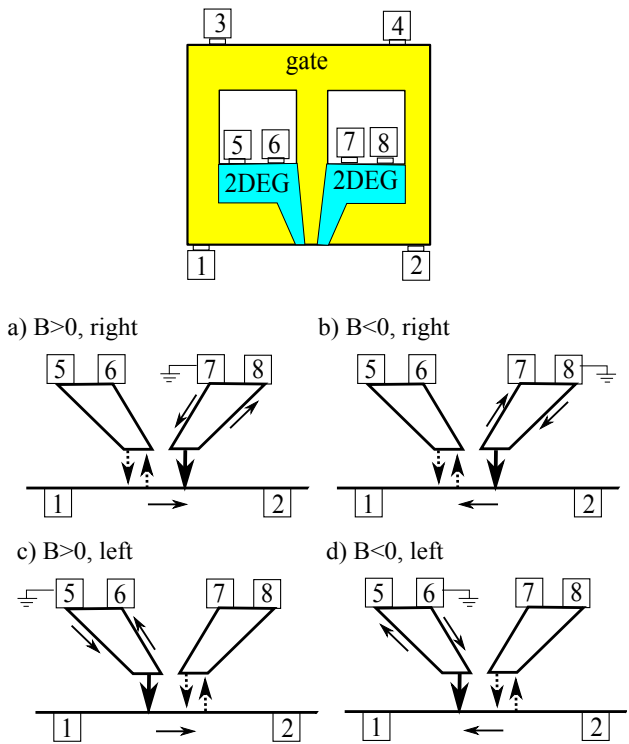}
\caption{(Color online)   Schematic diagram of the sample (not to scale). ES appear at the edges of etched regions (white), also at the border between the gate (yellow) and the uncovered 2DEG (light green). Ohmic contacts are denoted by bars with numbers.  (a-d) Experimental configurations for two injector positions and two field directions. Thin arrows indicate the electron drift along ES. Thick arrow denotes current in the injector junction.  Dotted ones are for the equilibrium (forward and backward) transitions in the detector gate-gap.
\label{sample}}
\end{figure}

A novel sample design realizes the theoretically proposed scheme with independent injector and detector~\cite{rosenow}, see Fig.~\ref{sample}, (top). Each sample has two macroscopic ($\sim 0.5\times 0.5\mbox{mm}^2$) etched regions inside, separated by a distance of 300~$\mu$m.  A split-gate partially encircles the etched areas, leaving uncovered two $L=5$~$\mu$m wide gate-gap regions, separated by 30~$\mu$m,  at the outer mesa edge. Ohmic contacts are placed along the mesa edges.

In samples with a smooth edge profile, edge states~\cite{buttiker} (ES) are represented by  compressible strips of finite width~\cite{shklovsky}, located at the intersections of the Fermi level and filled Landau levels. Every ES is characterized by a definite electrochemical potential, which is constant along ES except for the regions of charge exchange~\cite{buttiker,shklovsky}. For a bulk filling factor $\nu=2$, there are two co-propagating ES in each gate-gap region, separated  by the incompressible strip with local filling factor $\nu_c=1$. We deplete the 2DEG under the gate  to the same filling factor $g=1$, so that the $\nu_c=1$ incompressible state fully separates the outer and the two inner edges. They are only connected by inter-ES transport in the gate-gap junctions.  The maximum junction resistance does not exceed $R \sim(h/e^2) l_{eq}/L \sim$ 3~MOhm, where $l_{eq}/L\sim 100$ is the ratio between the maximum charge equilibration length~\cite{mueller} $l_{eq}$ and the  gate-gap width $L$. Because of finite $R$, one can expect $\mu_{out}=\mu_{in}$  for the electrochemical potentials of two ES in the equilibrium.

In the present experiment, we enforce inter-ES transport in one gate-gap junction (injector), by applying a {\em dc current} between the outer contact, labeled as 3, and one of the inner contacts (the ground), see Fig.~\ref{sample}. It causes energy dissipation in the injector in the form of plasmons, non-equilibrium electrons or phonons.  The other gate-gap junction serves as a detector: the energy can be absorbed here by stimulating inter-ES transitions, which would disturb the equilibrium $\mu_{in}=\mu_{out}$ in the detector junction. We monitor the ES potentials by high-impedance electrometers connected to the Ohmic contacts in Fig.~\ref{sample}. 

In our setup with co-propagating ES, the normal magnetic field $B$ defines the propagation direction for the charged transport along the outer mesa edge. There are four possible experimental configurations, depicted in Fig.~\ref{sample} (a)-(d), which are labeled by the magnetic field sign ($B>0$ or $B<0$) and by the position (right or left) of the injector gate-gap junction. 

The measurements are performed at a temperature of 30~mK. The obtained results for energy transport are independent of the cooling cycle, despite the carrier density differs slightly in different coolings. Standard two-point magnetoresistance is employed to determine the carrier density and to verify the contact quality. Magnetocapacitance measurements are used to find the available filling factors $g$ under the gate.

 \begin{figure}
\includegraphics*[width=0.95\columnwidth]{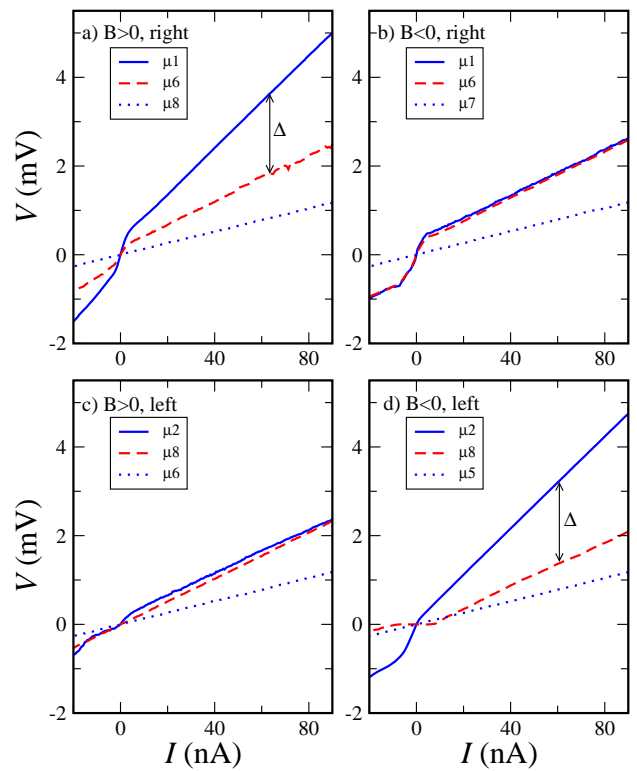}%
\caption{(Color online)Potentials $V_i$ of different Ohmic contacts ($eV_i=\mu_i$) vs. injector current for the experimental configurations depicted in Fig.~\protect\ref{sample}. Blue solid: potential $\mu_{out}$ of the outer ES in the detector, which is also expected for the inner one in the equilibrium $\mu_{out}=\mu_{in}$. Red dash:  measured potential of the inner ES $\mu_{in}$ in the detector. $\Delta$ denotes the difference $e\Delta=\mu_{out}-\mu_{in}$. Blue dots: the potential of the inner contact within the injector.   Positive $B=+3.72$~T and negative $B=-3.51$~T fields differ in value because of different coolings. Filling factors are $\nu=2, g=1$. \label{mu21}}
\end{figure}
 
\begin{figure}
\includegraphics*[width=0.95\columnwidth]{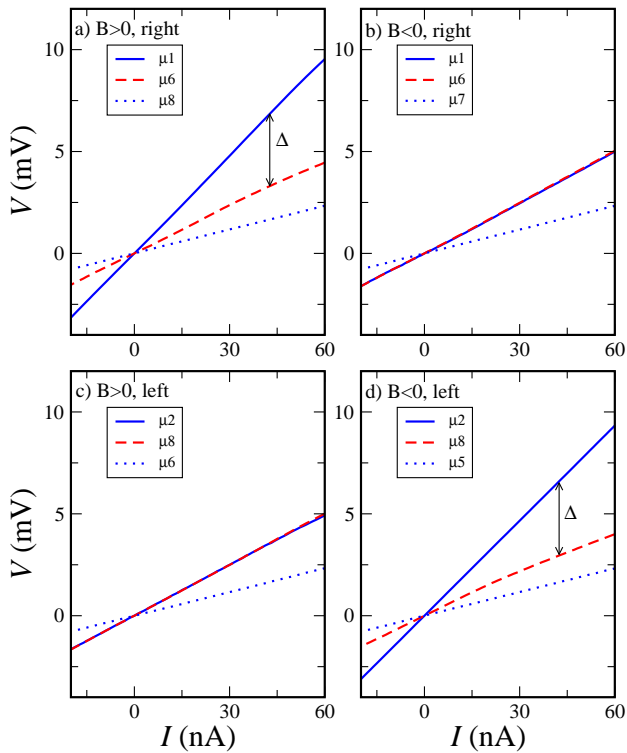}%
\caption{(Color online) Potentials $V_i$ of different Ohmic contacts ($eV_i=\mu_i$) vs. injector current in the fractional QH regime. The configurations are the same as in Fig.~\protect\ref{mu21}, the filling factors are $\nu=2/3, g=1/3$. Positive $B=+11.15$~T and negative $B=-10.53$~T fields differ in value because of different coolings. \label{mu2313}}
\end{figure}

The potentials of different Ohmic contacts are shown in  Fig.~\ref{mu21} for integer filling factors $\nu=2, g=1$ for all experimental configurations depicted in Fig.~\ref{sample}. The curves are obtained in a stationary regime, i.e. very slowly (about 3 hours per curve) to complete all the relaxation processes~\cite{ESreview}.

In the detector gate-gap junction, we indeed observe the equilibrium ES electrochemical potential distribution $\mu_{out}=\mu_{in}$ for two experimental configurations, see Fig~\ref{mu21} (b) and (c). The most surprising experimental finding is the fact that in two other cases there is a non-zero difference $e\Delta=(\mu_{out}-\mu_{in})$, see Fig.~\ref{mu21} (a) and (d). It occurs if the detector is situated upstream of the injector, {\em cp.} Fig.~\ref{sample} (a) and (d). The effect is present for both signs of the applied current. 

In our set-up, we not only know the direction of the electron drift in Fig~\ref{sample}, but can also obtain it  from the experimental curves~\cite{ESreview}. If the contact within the injector is situated so that  electrons reach it before the ground, its  potential reflects the charge transfer across ES within the injector~\cite{ESreview}. Since we apply a {\em current} through the injector junction, we do find this potential to be linear and independent on the experimental configuration  in a full current range, see Fig.~\ref{mu21} (blue dots). In contrast, the potentials of the outer contacts demonstrate a clear non-linear behavior, see Fig.~\ref{mu21},  because they are  sensitive to the transport regime within the injector (see ~\cite{ESreview} and discussion below).  Charge conservation for the injector junction demands an evident relation between electrochemical potentials~\cite{ESreview}, e.g. $\mu_2=\mu_1-2\mu_8$  for ($B>0, right$) configuration or $\mu_1=\mu_2-2\mu_7$ for ($B<0, right$) one for $\nu=2,g=1$. Since the proper relation is indeed fulfilled for any experimental configuration, we are sure about the current distribution in the sample: the transport current  only flows across the injector junction, while electrons drift from the contact 1 to the contact 2  for the field which we denote as positive $B>0$. 

Similar results as those in Fig.~\ref{mu21} are obtained for other bulk fillings with $\nu_c=1$, such as $\nu=3, g=1$ and $\nu=4/3, g=1$. Furthermore,  we observe the same effect also for transport across fractional $\nu_c=1/3$.  Fig.~\ref{mu2313} demonstrates finite $\Delta$ for the same two experimental configurations for $\nu_c=1/3$. The curves are linear even around zero bias, because of the smaller equilibration length~\cite{ESreview}. 

We can be sure that the detector is only connected with the injector through the edge. Zero $\Delta$ in Fig.~\ref{mu21} (b,c)  excludes a parasitic connection through the bulk, which would produce positive $\Delta$ of the same order for all four experimental configurations.  A similar effect would be produced by a parasitic ground within the detector region~\cite{appendix}. Fig.~\ref{mu2313} also confirms the observed effect for  an order of magnitude smaller detector resistance $R=6(h/e^2)$.

\begin{figure}
\includegraphics*[width=\columnwidth]{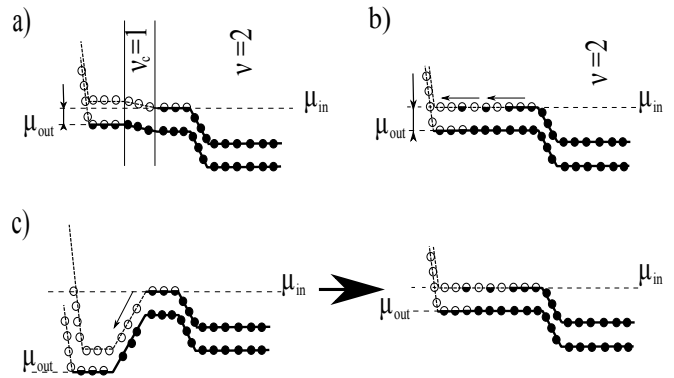}%
\caption{(Color online) Schematic diagrams of  the  energy levels in the gate-gap junction at integer filling factors $\nu=2, g=1$. Pinning of the Landau sublevels (solid) to the Fermi level (short-dash) is shown in the compressible regions at electrochemical potentials $\mu_{out}$ and $\mu_{in}$. Filled (half-filled) circles represent the fully (partially) occupied electron states. Open circles are for the empty ones. Arrows indicate electron transitions  at high imbalance.  (a)  Low imbalances $eV=\mu_{out}-\mu_{in}$ across the incompressible strip. (b) $eV$  reaches the spectral gap within $\nu_c=1$. (c) Evolution of higher imbalance along the gate-gap edge: applied in the corner of the injector gate-gap in Fig.~\protect\ref{sample}, it drops within 2-3~$\mu$m~\protect\cite{ESreview}. 
\label{band}}
\end{figure}

Since there is no parasitic connection between the injector and the detector, a finite $\Delta$ {\em in a stationary regime} implies that the equilibrium is {\em dynamic} within the detector junction~\cite{notthoff}. The 'forward' inter-ES transitions, which tend to equilibrate ES, should be compensated by counter 'backward' ones.  The necessary for $\nu_c=1$ electron spin flip is easily provided by  spin-orbit coupling~\cite{mueller} or by the flip-flop process~\cite{ESreview}, but the energy for backward transitions can only be transferred from the injector. Thus, Figs.~\ref{mu21},\ref{mu2313} demonstrate an energy transfer at the edge, which propagates counter to the electron drift.

This upstream energy transfer can only be provided by neutral excitations such as non-equilibrium phonons or neutral collective modes. To distinguish between these two, we start from the transport regimes~\cite{ESreview} across the injector gate-gap junction for integer $\nu_c$:

(i) Low imbalance: Here, the ES electrochemical potential imbalance is much smaller than the energy gap in the $\nu_c=1$ incompressible strip, see Fig~\ref{band} (a). This regime corresponds to a high-resistance $R \sim(h/e^2) l_{eq}/L$ observed for small $\mu_1, \mu_2$ in Fig.~\ref{mu21}. Two gate-gap junctions have different resistances because of their different real widths $L$.

(ii) High imbalance: When the electrochemical potential difference reaches the spectral gap (Fig.~\ref{band} (b)), electrons can overflow the initial potential barrier, see Fig.~\ref{band} (c), so the junction resistance is diminished. Some electrons are transferred elastically with farther relaxation outside the gate-gap junction, while others loss their energy within the junction.  This   regime corresponds to the linear behavior of $\mu_1$ and $\mu_2$  in Fig.~\ref{mu21} at high currents.

\begin{figure}
\includegraphics*[width=\columnwidth]{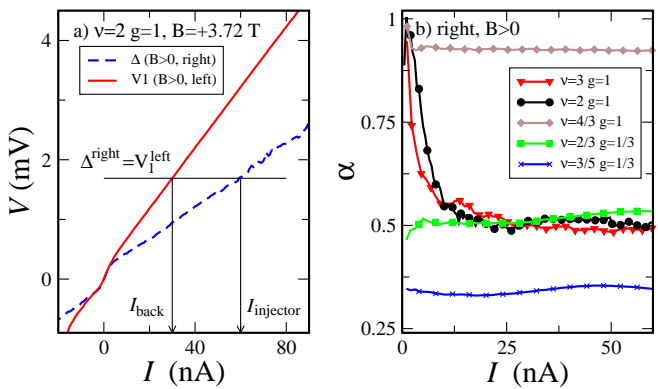}%
\caption{(Color online) (a) Potential $V_1=\mu_1^{left}/e$  in comparison with $\Delta^{right}$ for $B>0$, $\nu=2, g=1$. (b) Calculated $\alpha=I_{back}/I_{injector}$ (see the text) as a function of $I_{injector}$ for different filling factors, $B>0$.   \label{alpha}}
\end{figure}

To describe the efficiency for energy transfer, we define the ratio $\alpha=I_{back}/I_{injector}$ between the 'backward' current in the detector and the 'forward' current in the injector.  $I_{back}$ can be obtained from the measured $\Delta$ using the non-linear resistance of the detector gate-gap junction. This resistance is solely determined by the ES structure and the gate-gap width, so it can be obtained from the ES imbalance in the injector in a symmetric configuration. We therefore determine $\alpha$ as depicted in Fig.~\ref{alpha} (a). It is worth mentioning, that both curves in Fig.~\ref{alpha} (a) change their slopes at the same voltage, in contrast to, e.g., Fig.~\ref{mu21} (a). This is an additional argument that $\Delta$ originates from the non-linear resistance of the detector junction.

A value of $\alpha=1$ at low imbalances indicates a low dissipation of energy between the injector and the detector. If the transfer mechanism is the same, at high imbalances $\alpha$ should reflect a part of non-elastic inter-ES transitions in the injector, which is confirmed by data in Fig.~\ref{alpha} (b). The data coincide for the filling factors $\nu=3, g=1$ and $\nu=2, g=1$ since the involved ES are separated by the same $\nu_c=1$. Much higher $\alpha$ for $\nu=4/3, g=1$  reflects the fact that efficient elastic transitions are not reachable for the bulk $\nu=4/3$~\cite{highfrac}.  For $\nu_c=1/3$ (at $\nu=2/3$ and 3/5) $\alpha$ is practically independent of the injector current, but differs in a value possibly because of different structure of the bulk ground state. In this regime the linearity of the curves in Fig.~\ref{mu2313} confirms~\cite{ESreview} the presence of the gap at $\nu_c=1/3$, and therefore non-elastic transitions in the injector.   

It is very unlikely, that all  phonons emitted in the injector at low imbalances would be absorbed in the detector, resulting in $\alpha=1$. In contrast, collective modes are propagating along the edge, they are characterized by low dissipation, and their dispersion allows to transfer an appropriate energy~\cite{wen,rosenow,kane}. In our set-up it is a  dipole (neutral) collective excitation which is created by an intra-edge electron transition across the $\nu_c=g$ incompressible strip, see Fig.~\ref{band} (c). The neutral mode can propagate upstream along the low-density edge of the $\nu_c=1, 1/3$ incompressible strip if the density profile is reconstructed at the edge~\cite{yang}.

In summary, we experimentally observe an upstream energy transfer with respect to the electron drift  for the filling factors 1 and 1/3, which seems to be provided by the neutral collective mode at the reconstructed sample edge~\cite{yang}. The excitation of this mode is especially efficient in the overflowing process depicted in Fig.~\ref{band} (c), so the regime of high imbalance opens a direct access to the neutral mode. This is the keynote difference of the present experiment from Ref.~\cite{granger}, where a weak electron tunneling to the QH edge should mostly excite a fundamental (charged) mode.

We wish to thank  V.T.~Dolgopolov, D.E. Feldman, D.G. Polyakov, D. A. Bagrets, and B. I. Halperin for fruitful discussions. We gratefully acknowledge financial support by the RFBR, RAS, the Program ``The State Support of Leading Scientific Schools''.


\begin{thebibliography}{99}
\bibitem{wen} Xiao-Gang Wen, Phys. Rev. B \textbf{41}, 12838 (1990).
\bibitem{kane} C. L. Kane, M. P. A. Fisher,  and Polchinski, Phys. Rev. Lett. 72, 4129 (1994);  D. E. Feldman, F. Li, Phys. Rev. B 78, 161304-161307 (2008); E. Grosfeld, S. Das, Phys. Rev. Lett. 102, 106403 (2009); B. Rosenow and B. I. Halperin, Phys. Rev. B 81, 165313 (2010).
\bibitem{granger} G. Granger, J. P.  Eisenstein, J. L. Reno,  Phys. Rev. Lett. 102, 086803
(2009).
\bibitem{heiblum} Aveek Bid, Nissim Ofek, Hiroyuki Inoue, Moty Heiblum, Charles Kane, Vladimir Umansky, Diana Mahalu, Nature 466, 585, (2010).
\bibitem{yang} Xin Wan, E. H. Rezayi, and Kun Yang, Phys. Rev. B 68, 125307 (2003); C. d. C. Chamon and X. G. Wen, Phys. Rev. B \textbf{49}, 8227 (1994).
\bibitem{eisen} J. P. Eisenstein, L. N. Pfeiffer, and K. W. West, Phys. Rev. Lett. 68, 674 (1992).
\bibitem{rosenow} S. Takei, M. Milletarì, and B. Rosenow,  Phys. Rev. B \textbf{82}, 041306(R) (2010).
\bibitem{buttiker} M. B\"uttiker, Phys. Rev. B {\bf 38}, 9375 (1988).
\bibitem{shklovsky} D. B. Chklovskii, B. I. Shklovskii, and L. I. Glazman, Phys. Rev. B {\bf 46}, 4026 (1992).
\bibitem{ESreview} A. W\"urtz, et. al., Phys. Rev. B {\bf 65}, 075303 (2002). For a review see E. V. Deviatov, A Lorke, phys. stat. sol. (b) 245, 366 (2008).
\bibitem{mueller}G. M\"uller, et. al.,  Phys. Rev. B {\bf 45}, 3932 (1992).
\bibitem{appendix} Even an unbelievable mistake in relative positions of the gate-gap regions would result in antisymmetric $\Delta$ with respect to the magnetic field sign. 
\bibitem{notthoff} C. Notthoff, K. Rachor, D. Heitmann, and A. Lorke,  Phys. Rev. B {\bf 80}, 205320 (2009).
\bibitem{highfrac}E. V. Deviatov, A. Lorke, and W. Wegscheider, Phys. Rev. B 78, 035310 (2008).

\end{thebibliography}
 \end{document}